\documentclass[aps,prl,twocolumn,superscriptaddress]{revtex4}
\usepackage{
amsfonts,amssymb,epsfig,amsmath,graphics,subfig}
\usepackage{color}
\usepackage{hyperref,graphicx}
\usepackage{cancel}

\renewcommand{\text}[1]{#1}

\newcommand{\be}{\begin{equation}}
\newcommand{\ee}{\end{equation}}
\newcommand{\ben}{\begin{displaymath}}
\newcommand{\een}{\end{displaymath}}
\newcommand{\bea}{\begin{eqnarray}}
\newcommand{\eea}{\end{eqnarray}}
\newcommand{\bean}{\begin{eqnarray*}}
\newcommand{\eean}{\end{eqnarray*}}
\newcommand{\nn}{\nonumber \\}
\newcommand{\ba}{\begin{array}}
\newcommand{\ea}{\end{array}}
\newcommand{\bi}{\begin{itemize}}
\newcommand{\ei}{\end{itemize}}

\newcommand{\reef}[1]{(\ref{#1})}

\def\dblone{\hbox{$1\hskip -1.2pt\vrule depth 0pt height 1.6ex width 0.7pt
                  \vrule depth 0pt height 0.3pt width 0.12em$}}

\def\tdgamma{\gamma}
\def\tdrho{\gamma}

\newcommand{\bbR}{{\mathbb{R}}}

\begin{document}
\title{Universal fermionic spectral functions from string theory}

\author{Jerome P. Gauntlett}
\affiliation{Theoretical Physics Group, Blackett Laboratory,
  Imperial College, London SW7 2AZ, U.K.}
\author{Julian Sonner} 
\affiliation{Theoretical Physics Group, Blackett Laboratory,
  Imperial College, London SW7 2AZ, U.K.}
\affiliation{D.A.M.T.P. University of Cambridge,
 Cambridge, CB3 0WA, U.K.}
\author{Daniel Waldram}
\affiliation{Theoretical Physics Group, Blackett Laboratory,
  Imperial College, London SW7 2AZ, U.K.}
\begin{abstract}
We carry out the first holographic calculation of a fermionic response
function for a strongly coupled $d=3$ system with an explicit 
$D=10$ or $D=11$ supergravity dual. By considering
the supersymmetry current, we obtain a universal result
applicable to all $d=3$ $N=2$ SCFTs with such duals. 
Surprisingly, the spectral function does not exhibit a Fermi surface,
despite the fact that the system is at finite charge density. We show
that it has a phonino pole and at low frequencies there is a depletion
of spectral weight with a power-law scaling which is governed by a
locally quantum critical point.
\end{abstract} 

\maketitle

\setcounter{equation}{0}


\section{Introduction}

The AdS/CFT correspondence provides a powerful framework for studying
strongly coupled quantum field theories and 
has recently been used as a theoretical laboratory for studying 
condensed matter systems. Some of the most interesting
``AdS/CMT'' studies have focussed on calculating fermionic 
response functions, not least
in the hope of obtaining a better understanding of the deeply vexing
non-fermi liquids that are seen in a variety of materials at finite
charge density including the heavy fermion and high-$T_c$ cuprate
superconductors.  

The ground-breaking works
\cite{Lee:2008xf,Liu:2009dm,Cubrovic:2009ye,Faulkner:2009wj} 
gave the first such ``holographic'' calculations of fermion spectral
functions using phenomenological or ``bottom-up'' models. The AdS/CFT
correspondence states that certain classes of field
theories have specific dual gravitational descriptions, determined by
their realisation in string theory. In the bottom-up approach, rather
than identify a specific string dual, one simply postulates a
particular theory of gravity with some simple matter content and
couplings and assumes that it captures the essential features of 
potential dual field theories. Specifically the original papers
considered the Dirac equation for a minimally coupled spin 1/2 fermion
with mass $m$ and charge $q$ in the 
gravitational background of a four-dimensional
AdS-Reissner-Nordstr\"om ($AdS_4$-RN)  
black-brane. With appropriate boundary conditions, this encodes a
fermionic response function at finite temperature and chemical
potential. 
It was shown that the resulting spectral function can exhibit a Fermi
surface with non-Fermi liquid scaling for certain values of $m$ and
$q$. It can also have an interesting oscillatory behaviour, periodic
in the logarithm of the frequency. While the existence of these Fermi
surfaces depend both on the full 
$AdS_4$-RN geometry,  
it was shown in \cite{Faulkner:2009wj} (see also
\cite{Faulkner:2010tq}), how their low-frequency scaling behvaiour can
be beautifully understood as a consequence of the $AdS_2\times\bbR^2$
``IR'' region of the spacetime that is dual to an emergent
one-dimensional conformal field theory (CFT).  

The validity of phenomenological models rests on the hope that either
somewhere in the landscape of string theory backgrounds the model will
be realised exactly,
and hence the holographic calculations relate to a specific dual field
theory, or alternatively, the gravitational model may only be realised
approximately but the features are sufficiently 
robust to capture properties of some actual field theory. 
Although significantly more difficult, it is
clearly essential to study ``top-down'' models
in which one is carrying out holographic calculations within a
explicit string theory setting and hence obtaining 
results for  {\it bona fide} dual field theories.

The purpose of this letter is to communicate the first such calculations of
fermion spectral functions in ten- or eleven-dimensional supergravity,
the low-energy limit of string/M-theory. The most robust and controlled examples of holography
are for supersymmetric conformal field theories (SCFTs) and we will restrict our considerations to this class. 
Remarkably, as we will
explain, our results will be valid not just for a single field theory
but for an {\it infinite} number. 

We analyse the response function of 
the {\it universal} 
spin-$\frac{3}{2}$ 
supersymmetry current, or ``supercurrent",  in the {\it general}
infinite class of $d=3$, $N=2$ SCFTs that have
dual gravitational backgrounds of the form $AdS_4\times M$ in either
$D=10$ or $D=11$ supergravity. The 
supercurrent, the energy-momentum tensor and the global abelian
$R$-symmetry current of the SCFT comprise a supermultiplet.  
It is possible to isolate this universal sector from all other
operators because, from the gravitational point of view, given a
Kaluza-Klein (KK) reduction of $D=10$ or $D=11$ supergravity on any
appropriate manifold $M$ one can then consistently truncate an
infinite tower of fields leaving minimal $N=2$ $D=4$ gauged supergravity
\cite{Gauntlett:2007ma}. The field content of this gauged supergravity
consists of a metric, a gauge field and a Dirac gravitino, which are
precisely dual to the energy-momentum tensor, the global abelian
$R$-symmetry current and the fermionic supercurrent of the SCFT,
respectively.

We consider the electrically charged $AdS_4$-RN black-brane solution
which provides the dual description of the SCFTs at finite temperature
$T$ and chemical potential $\mu$ with respect to the global
$R$-symmetry, both of which break the supersymmetry.
It is possible that the SCFT undergoes a phase transition at some
critical temperature $T_c$, which will involve other KK fields, and if
it does then the $AdS_4$-RN description will be valid only for
temperatures above $T_c$. It is an open question whether or not there
are SCFTs which do not have such phase transitions and hence are 
described by the extremal $AdS_4$-RN black-brane all the way down to $T=0$.

We calculate the supercurrent response function 
by solving the
linearised gravitino equations in the $AdS_4$-RN background, 
as a function of frequency $\omega$ and momentum $k\equiv |{\bf k}|$. 
We find that there is no log-periodic behaviour, in contrast to the
bottom-up model results. 
Furthermore, 
it does {\it not} have a Fermi surface,
i.e. a quasi-particle pole with $\omega=0$ and $k\ne 0$, as one might
have expected for matter at finite charge density
\cite{Huijse:2011hp}. This surprising result underscores the
importance of the top-down approach. Further study will be required to
determine whether a Fermi surface will be seen in different
response functions or whether they are in fact absent in
these holographic theories.  

The spectral function has other interesting features. It
has a ``phonino pole"
\cite{Lebedev:1989rz,Kovtun:2003vj,Policastro:2008cx} located
at $\omega+\mu=0$ and $k=0$, reflecting the broken supersymmetry.
We also find a depletion of spectral weight at low frequencies,
as seen in \cite{Edalati:2010ww,Guarrera:2011my}, where bulk dipole couplings
were considered in a bottom-up context. In \cite{Edalati:2010ww} this
behaviour was interpreted as a dynamical gap dual to something akin to
a Mott insulator. A subsequent discussion of this interpretation can be found in \cite{Guarrera:2011my}. Here we will show that at zero temperature
the spectral function vanishes when $\omega=0$. Furthermore, the
low-frequency behaviour is weakly gapped (and thus unlike a Mott gap) and
determined by an emergent one-dimensional, ``locally quantum
critical", CFT, dual to the IR $AdS_2\times\bbR^2$ part of the
geometry. This behaviour persists, albeit in a softened way, for 
non-zero temperatures. 

In \cite{companion} 
we present more details of the rather technical calculations as well
as some additional results.

\section{Supercurrent response function}
\label{sec:corrfn}
Let $S_{\alpha}$ be the conserved supercurrent operator of the 
$d=3$ SCFT. 
It is a complex vector-spinor, where $\alpha$ is the vector index, has
conformal dimension $\Delta=\frac{5}{2}$, 
and is charged under the global $R$-symmetry. 
We will calculate the retarded correlation function 
$G_{{\alpha}{\beta}}(p)=
\big<S_{\alpha}(p)\bar{S}_{\beta}(0)\big>_{Ret}$
at finite temperature and chemical potential,
exploiting the fact that the expectation value of the supercurrent 
in the presence of a vector-spinor source $a_\alpha$, at linearised
order,  
is given by
\begin{align}\label{form}
\langle S_{\alpha}\rangle= iG_{\alpha\beta}a^\beta\, .
\end{align}
The supercurrent is conserved and, because we have an SCFT, gamma-traceless: 
$p^{\alpha}\langle
S_{\alpha}\rangle={\tdgamma}^{{\alpha}}\langle S_{\alpha}\rangle =    0\,$ 
where $\tdgamma^{{\alpha}}$ are $d=3$ gamma-matrices. 
Since we are considering the SCFT at finite $\mu$, which can be viewed
as weakly gauging the $R$-symmetry, we have $p^{\alpha} =
(\tilde\omega,{\bf k})$ with $\tilde \omega\equiv \omega+\mu\,$. 
The source can be taken to satisfy 
\begin{align}\label{sceconds}
\gamma^\alpha a_\alpha=0,\qquad 
\delta a_\alpha=\left(\delta_\alpha^\beta-\tfrac{1}{3}\gamma_\alpha \gamma^\beta
   \right)p_\beta\epsilon\, ,
\end{align}
where the second equation arises from the weak gauging of the
supersymmetry. Of course the supercurrent itself, and hence its
expectation value, is gauge invariant. 

The four independent components of $G_{\alpha\beta}$ can be extracted by
introducing a basis of 3d vector-spinors $e^{(i)}_{\alpha}$, $i=1,2$, 
satisfying $
\tdrho^{\alpha}e^{(i)}_{\alpha}=p^{\alpha}e^{(i)}_{\alpha}=0$ and the
normalisation condition 
$\bar{e}^{(i)}_\alpha e^{(j)\alpha}=-2p^2\epsilon^{(i)(j)}$. We can then write $
   G_{\alpha\beta} 
      = t_{ij}  e^{(i)}_{\alpha}\bar{ {e}}^{(j)}_{\beta}\,$,
where the $t_{ij}$ 
are the four independent components of $G_{\alpha\beta}$. The
$d=3$ SCFT is invariant under spatial rotations and parity. We can use this 
to choose $p^{\alpha}=(\tilde\omega,k,0)$, where 
$k\equiv|{\bf k}|$, and show that 
$t_{12}=t_{21}=0$ and 
$t_{22}(\omega,k)=t_{11}(\omega,-k)$. Thus the correlation 
function is determined by a single function $t_{11}$. 
Our objective is to calculate $t_{11}(\omega,k)$, and more specifically 
the spectral function, $A(\omega,k) $, defined as 
\be
A(\omega,k) \equiv {\rm Im}\, \,t_{11}(\omega,k)\, .
\ee

\section{Holographic calculation} 

\subsection{$N=2$ gauged supergravity in $D=4$}\label{n2gs}
The field content of minimal $N=2$ gauged supergravity in $D=4$ 
\cite{Freedman:1976aw,Fradkin:1976xz}
consists of a metric $g_{\mu\nu}$, a gauge field ${\cal A}_\mu$ and a single Dirac
gravitino $\psi_\mu$. This theory admits the $AdS_4$-RN black-brane solution given by
\begin{equation}\label{adsrn1}
   d s^2 = - fdt^2 + \frac{dr^2}{f}  
     + \frac{r^2}{\ell^2}\left( dx^2 + dy^2 \right) \,,\qquad {\cal A}=\phi dt\,,
\end{equation}
with
\begin{equation}
\begin{aligned}\label{adsrn2}
 f&=\frac{r^2}{\ell^2}-\frac{r_+}{r}\left(\frac{r_+^2}{\ell^2}+\ell^2\mu^2\right)+\ell^2\mu^2\frac{r_+^2}{r^2}\\
  \phi &= \mu\ell \left( 1-\frac{r_+}{r} \right) .
\end{aligned}
\end{equation}
The location of the horizon is $r=r_+$.
The temperature of the black-brane is given by $T=(3r_+/\ell^2-\ell^2\mu^2/r_+)/4\pi$.
When $T=0$, as $r\to r_+$ the black-brane solution approaches $AdS_2\times \mathbb{R}^2$
with the radius of the $AdS_2$ given by $L_{(2)}=\ell/\sqrt{6}$. 

We will study the equation of motion of the gravitino at the linearised level in the $AdS_4$-RN background \eqref{adsrn1},\eqref{adsrn2}.
One can use the local superysmmetry to fix the gauge
$D^\mu \psi_\mu = \Gamma^\mu \psi_\mu =0$, where $\Gamma^\mu$ are $D=4$ gamma-matrices, 
and we then obtain 
\bea\label{graveq}
 \left( \cancel{D} -m \dblone - \tfrac{1}{2}i{\cal F}^{\mu\nu}\Gamma_{\mu\nu} \right)\psi_\rho +i{\cal F}^{\mu\nu}\Gamma_{\mu}\Gamma_\rho \psi_\nu=0\, ,
\eea
where $D\equiv \nabla - i q{\cal A}$, ${\cal F}=d{\cal A}$ and $ q=-m =\frac{1}{\ell}\,$.
There are residual gauge transformations, which we fix later.
We note the presence of Pauli terms.

\subsection{Asymptotic behaviour of solutions}

Under the AdS/CFT correspondence, 
to calculate $G_{\alpha\beta}$ one solves
the linearised equations of motion for the dual gravitino field in
$AdS_4$-RN, imposing ingoing boundary conditions at the horizon, and
studies the asymptotic expansion of the field as $r\to\infty$. This describes
a linearised perturbation of the CFT, where the $r^{\Delta-3}$ term
encodes the source $a_\alpha$ and the $r^{-\Delta}$ term encodes
the resulting expectation value of the operator $\langle S_{\alpha}\rangle$.

We will assume throughout that the time and space dependence of the
gravitino is given by $e^{-i\omega t+i{\bf k}\cdot{\bf x}}$. 
As $r\to\infty$, schematically, we have 
\begin{align}\label{expPsi}
\psi &= r^{-1/2}\psi_{-1/2}+r^{-3/2}\psi_{-3/2}+r^{-5/2}\psi_{-5/2}\nn
&\qquad +r^{-7/2}\psi_{-7/2}+r^{-7/2}\log r\phi_{-7/2}+\dots\, ,
\end{align}
where $\psi_{-1/2}$ etc are functions of three-momentum
$p^{\hat\alpha}=(\tilde \omega,k,0)$. There is an analogous expansion
for the residual gauge transformations, 
fixed by two parameters $\varepsilon_{1/2}$ and $\varepsilon_{-7/2}$
appearing at orders $r^{1/2}$ and $r^{-7/2}$
respectively. 

It is then natural to decompose all these components under the
asymptotic $d=3$ Lorentz symmetry that appears as $r\to\infty$.  
Using that gauge conditions and residual gauge transformations, one
finds that the solution
is completely determined by a pair of $d=3$ vector-spinors $a_\alpha$
and $b_\alpha$ satisfying 
\begin{align}\label{condsonanadb}
\gamma^\alpha a_\alpha =0,\qquad 
\delta a_\alpha=\left(\delta_\alpha^\beta-\tfrac{1}{3}\gamma_\alpha
   \gamma^\beta\right)p_\beta\epsilon \, , \nn
\gamma^\alpha b_\alpha=0,\qquad p^\alpha b_\alpha=0,\qquad \delta b_\alpha=0\, ,
\end{align}
where $\epsilon$ is a $d=3$ spinor that determines
$\varepsilon_{1/2}$. To order $r^{-3/2}$ the expansion is completely
determined by $a_\alpha$. Terms in $a_{\alpha}$ also appear at order
$r^{-5/2}$ and this leads to some ambiguity in defining the new
independent data 
$b_{\alpha}$, that appears at this order. 
However, it can be fixed uniquely using the second set
of conditions in~\eqref{condsonanadb}. 
The full expansion 
requires the introduction of another spinor in $\psi_{-7/2}$, but this
data can be gauged away using $\varepsilon_{-7/2}$.

Since the supercurrent is a $\Delta=5/2$ operator in the dual $d=3$
SCFT, the source is fixed by the $r^{-1/2}$ expansion data and the expectation value by the $r^{-5/2}$ expansion data, and hence can
be identified with $a_\alpha$ and $b_\alpha$, respectively.
Furthermore, \eqref{form} allows us to write $b_\alpha=iG_{\alpha\beta} a^\beta$
and we can show that
\begin{align}\label{teeoneone}
t_{11} = - \frac{i\bar{e}^{(2)}_\alpha b^\alpha}{2p^2\,\bar e^{(1)}_\beta
  a^\beta}\, .
\end{align}
This is invariant under residual gauge transformations.

\subsection{Solving the gravitino equation}\label{sec:gravitinoeq}

A convenient way to solve the gravitino equation \eqref{graveq} in the
$AdS_4$-RN background and impose the ingoing boundary conditions,  
is to dimensionally reduce on the two spatial directions $x,y$ and
decompose into $Spin(1,1)$ representations. Subject to the gauge
conditions $\Gamma^\mu\psi_\mu=D^\mu\psi_\mu=0$ there are 8
independent complex components in $\psi_\mu$. After dimensional
reduction these can be written in terms of functions of $r$ labeled
$u^{(s)}$ and $v^{(s)}$, where
$s=-\frac{3}{2},-\frac{1}{2},\frac{1}{2},\frac{3}{2}$ refer to the
helicity of the $Spin(1,1)$ representation. The two sets of functions,
$u^{(s)}$ and $v^{(s)}$, are parity eigenstates and map into each
other under a rotation by $\pi$ in the $x,y$ plane.

Using this decomposition 
the gravitino equations \reef{graveq}
in the $AdS_4$-RN background \reef{adsrn1} are 
equivalent to a system of linear ODEs for $u^{(s)}$ and 
$v^{(s)}$. The parity and rotational symmetries imply that the ODEs
for $u^{(s)}$ and $v^{(s)}$ do not mix and map into to each other 
if one replaces $k\to -k$. 
At finite temperature the horizon at $r=r_+$ is a regular singular
point of the ODEs and we can develop a Frobenius expansion. Writing
$\delta_s(\omega) = s + \frac{i\omega}{2\pi T}\,$ 
the solutions have leading-order behaviour
\bea
u^{(s)} &=& (r-r_+)^{-\frac{\delta_s(\omega)}{2}} u^{(s)}_0  + \dots\,,\qquad s = \tfrac{3}{2},-\tfrac{1}{2}\, ,\nn
u^{(s)} &=& (r-r_+)^{\frac{\delta_s(\omega)}{2}}u^{(s)}_0  +\dots\,,\qquad s = \tfrac{1}{2}\,,-\tfrac{3}{2}\, ,
\eea
where the $u^{(s)}_0$ are four arbitrary complex coefficients. Thus
the ingoing boundary condition at the horizon is given by
$u^{(1/2)}_0$=$u^{(-3/2)}_0$=$0$.
The residual gauge transformations allows us to gauge away either
$u^{(-1/2)}_0$ or  $u^{(3/2)}_0$. The situation for $T=0$ is slightly
more subtle and is explained in \cite{companion}.
It can be shown that if $(u^{(-s)})^*$ is a solution then so is
$u^{(s)}$. This is related to the action of 
time reversal and will be useful below.

Having solved the gravitino equations in this way,
one can rewrite $\psi_\mu$ in terms of $u^{(s)}$ and $v^{(s)}$. 
compare with the  asymptotic expansion, and obtain the boundary data
$a_{\alpha}$ and $b_{\alpha}$.
We then obtain $t_{11}$ from \eqref{teeoneone}.

\section{Results}
\label{results}
We now summarise some results for the spectral
function based on solving the ODEs numerically. 

\begin{figure}[t!]
\includegraphics[width=0.38\textwidth]{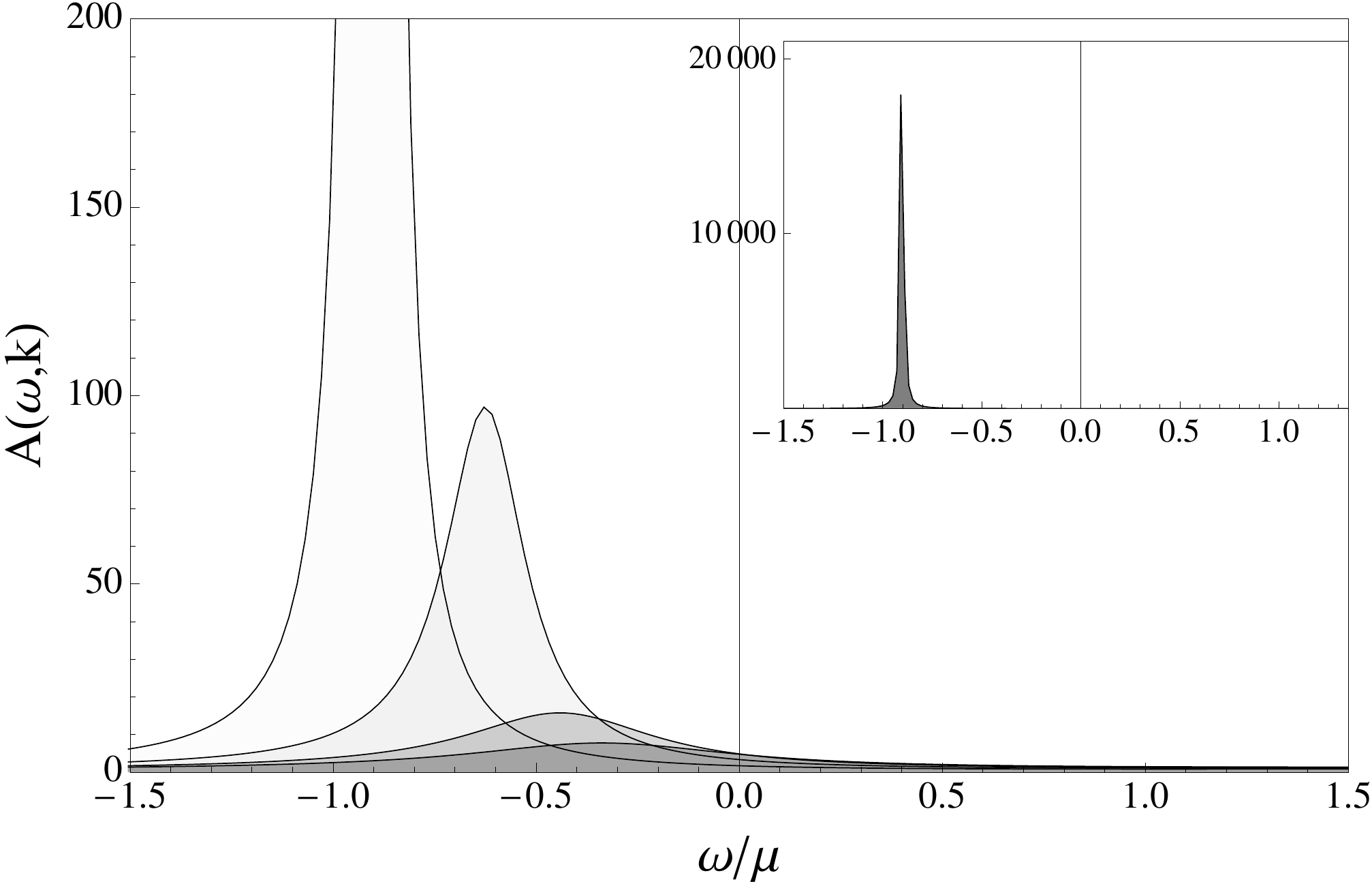}\hskip1em \includegraphics[width=0.38\textwidth]{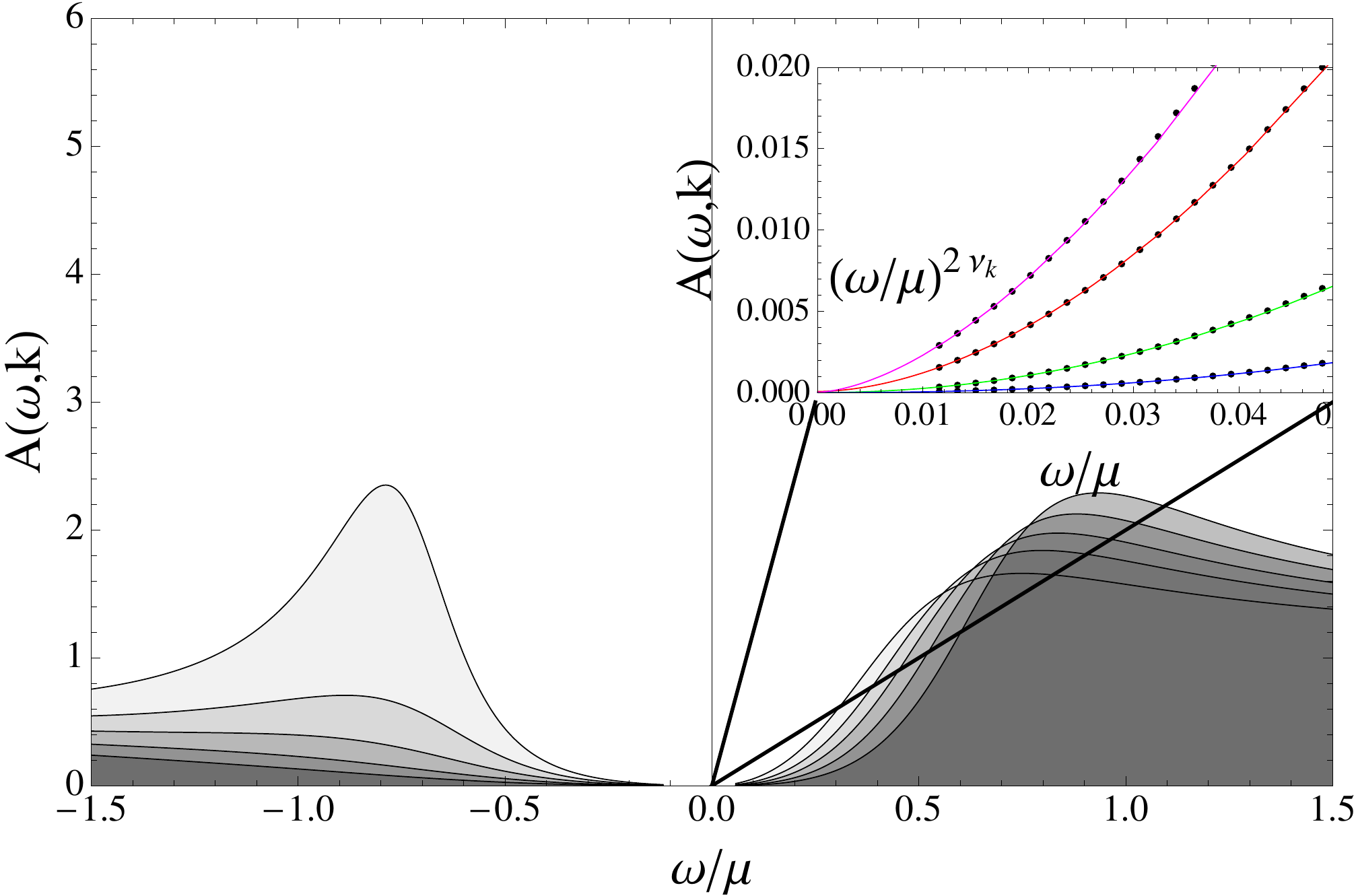}
\caption{The spectral function $A(\omega,k)$ 
The top panel is for $T/\mu=0.44$ and momenta $k\ell\in (0.1,1.1)$,
with larger values of $k$ in darker shades of grey. The bottom panel
is for $T=0$ and $k\ell\in (1.2,2.1)$.
\label{fig:Spectrum_1}}
\end{figure}

As illustrated in figure \ref{fig:Spectrum_1}, the most prominent
feature 
for $T\ne 0$ is the large peak near
$\tilde\omega=0$ associated with a pole of $G_{\alpha\beta}$ at $(\tilde\omega,k)= (0,0)$. This long-wavelength Goldstino 
pole has been discussed before in a hydrodynamical context.
In supersymmetric theories in addition to ordinary sound waves 
there are weakly damped
propagating ``super-sound'' waves, or ``phoninos''
\cite{Lebedev:1989rz,Kovtun:2003vj}.    
At $\mu=0$, this gives a pole in $G_{\alpha\beta}$ 
at $(\omega,k)=(0,0)$~\cite{Kovtun:2003vj},
and was analysed holographically in \cite{Policastro:2008cx}.

In our case $\mu\ne 0$, and the 
weakly gauged $R$-symmetry means 
the phonino 
pole is shifted to 
$(\tilde\omega,k)=(0,0)$, exactly as in 
figure \ref{fig:Spectrum_1}.
For higher values of $k$, this peak disappears. At the same time the spectral weight gets redistributed to positive $\omega$, where a bump
develops. 
At low temperatures and small $\omega$ there is a region of the order
of the chemical potential, where the density of states is depleted. An
analogous feature was interpreted in \cite{Edalati:2010ww} as a hard
(Mott) gap.

Some results for the spectral function for $T=0$ are shown in
figure \ref{fig:Spectrum_1}. The phonino pole is still present at
$(\tilde\omega,k)=(0,0)$, much as in the top panel. We also see that the spectral function
vanishes at $\omega=0$ for all values of $k$. In fact there is a scaling of the form $A\propto
\omega^{2\nu_k}$ corresponding to a soft power-law gap.

We can derive this behaviour analytically.
Indeed, by the method of matched asymptotic expansions, as in \cite{Faulkner:2009wj}, at $T=0$ and at leading order in $\omega$
we can show
\be
t_{11} (\omega,k)= t_{11}(0,k) \left( 1 + C(k) {\cal G}(\omega,\nu_k) + \cdots\right)\, ,
\ee
where
\bea
{\cal G}(\omega,\nu_k) &=& e^{-i\pi\nu_k}\frac{\Gamma(-2\nu_k)}{\Gamma(2\nu_k)}
 \frac{\Gamma(-1 -  \frac{i}{2\sqrt 3}     +\nu_k)}{\Gamma(-1 -  \frac{i}{2\sqrt 3} - \nu_k)} 
 \left( 2\omega L_{(2)} \right)^{2\nu_k}\nn
\eea
with $\nu_k =\sqrt{   \frac{7}{12} +\frac{k^2}{2\mu^2}}\,$. The function $C(k)$ is independent of $\omega$ and depends on the UV data of the system.
Note that since $\nu_k$ is real, for any $k$, there is no 
periodic log oscilliatory
 behaviour as seen in the bottom-up models.

If $t_{11}(0,k)$ is real then we can immediately extract the scaling relation for the spectral function
\be
A(\omega,k) \propto \omega^{2\nu_k}\, ,
\ee
for small $\omega$, exactly as we see in our numerical results.
The reality of $t_{11}(0,k)$ follows 
from the $u^{(s)}\to (u^{(-s)})^*$ symmetry we mentioned above. 
Thus the vanishing of the spectral weight at $\omega=0$ and $T=0$ is not a hard Mott-like  
gap but rather 
a power-law, characteristic of a local massless sector of states associated with the $AdS_2$ factor of the bulk near-horizon region.

Finally, it would be worthwhile to extend the results of this paper and \cite{companion}
to study fermion spectral functions in the more involved top-down models of \cite{Gauntlett:2009zw}. 
This would be particularly interesting as they include non-supersymmetric CFTs whose gravity duals are known
to be perturbatively stable. Furthermore,
it would be interesting to elucidate the impact of the superfluid phase 
\cite{Gauntlett:2009dn,Gauntlett:2009bh} at low temperatures and also
to see whether the additional bulk fermions reveal any underlying Fermi surface.

\subsection*{Acknowledgements}
We would like to thank J. Bhaseen, A. Buchel, A. Green, S. Hartnoll, J. Laia, J. McGreevy, R. Myers, S. Sachdev,
K. Schalm, D. Tong, D. Vegh, C. Warnick and J. Zaanen for helpful discussions.
JPG is supported by an EPSRC Senior Fellowship and a Royal Society Wolfson Award.
JS is supported by EPSRC and Trinity College Cambridge and thanks the GGI, Firenze and, along with JPG, the Aspen Center for Physics for hospitality during this work.

\end{document}